\documentclass[lettersize,journal]{IEEEtran}
\usepackage{amsmath,amsfonts}
\usepackage{algorithmic}
\usepackage{array}
\usepackage[caption=false,font=normalsize,labelfont=sf,textfont=sf]{subfig}
\usepackage{textcomp}
\usepackage{stfloats}
\usepackage{url}
\usepackage{verbatim}
\usepackage{graphicx}
\usepackage{bm}
\usepackage{algorithm}
\usepackage{algorithmic}
\def\BibTeX{{\rm B\kern-.05em{\sc i\kern-.025em b}\kern-.08em
    T\kern-.1667em\lower.7ex\hbox{E}\kern-.125emX}}
\usepackage{balance}
\begin{document}
\title{Improved AFSA-Based Beam Training \\ Without CSI for RIS-Assisted ISAC Systems}

\author{
	\IEEEauthorblockN{Yunxiang Shi\IEEEauthorrefmark{1}, Lixin Li\IEEEauthorrefmark{1}, Wensheng Lin\IEEEauthorrefmark{1}, Wei Liang\IEEEauthorrefmark{1}, and Zhu Han\IEEEauthorrefmark{2}}
	
	\IEEEauthorblockA{\IEEEauthorrefmark{1}School of Electronics and Information, Northwestern Polytechnical University,
		Xi'an, Shaanxi 710129, China}
	
	\IEEEauthorblockA{\IEEEauthorrefmark{2}University of Houston, Houston, USA}
	
	Email:\ shiyunxiang@mail.nwpu.edu.cn,\{linwest, lilixin, liangwei\}@nwpu.edu.cn, zhan2@uh.edu
	
	\thanks{Corresponding authors: Lixin Li, Wensheng Lin.
		
		This work was supported in part by National Natural Science Foundation of China under Grant 62101450, in part by the Young Elite Scientists Sponsorship Program by the China Association for Science and Technology under Grant 2022QNRC001, in part by Aeronautical Science Foundation of China under Grants 2022Z021053001 and 2023Z071053007, in part by the Open Fund of Intelligent Control Laboratory, 
		in part by NSF CNS-2107216, CNS-2128368, CMMI-2222810, ECCS-2302469, US Department of Transportation, Toyota and Amazon.}
	
}

\maketitle

\begin{abstract}
In this paper, we consider transmit beamforming and reflection patterns design in reconfigurable intelligent surface (RIS)-assisted integrated sensing and communication (ISAC) systems, where the dual-function base station (DFBS) lacks channel state information (CSI). To address the high overhead of cascaded channel estimation, we propose an improved artificial fish swarm algorithm (AFSA) combined with a feedback-based joint active and passive beam training scheme. In this approach, we consider the interference caused by multipath user echo signals on target detection and propose a beamforming design method that balances both communication and sensing performance. Numerical simulations show that the proposed AFSA outperforms other optimization algorithms, particularly in its robustness against echo interference under different communication signal-to-noise ratio (SNR) constraints.
\end{abstract}


\section{Introduction}
\IEEEPARstart{I}{ntegrated} sensing and communication (ISAC) systems are envisioned to play a crucial role in 5th generation mobile communication technology (5G) and 6th generation mobile networks (6G) \cite{b1}, \cite{b2}. Unlike separate radar and communication systems, ISAC enables shared use of limited spectrum by integrating both sensing and communication in the same frequency band, reducing system costs while enhancing performance in both areas.

ISAC systems, which operate across a wide spectrum, including the THz band \cite{b1}, leverage MIMO beamforming to enhance performance \cite{b5}, while dual-function radar communication base stations (DFBS) enable the simultaneous integration of communication and sensing capabilities \cite{b6}. ISAC typically reuses communication waveforms, embeds communication symbols into radar waveforms, or designs hybrid waveforms with precoders to reduce interference \cite{b10}.

Millimeter wave (mmWave) frequencies offer high data rates, large bandwidth, and precise sensing, but suffer significant path loss, especially in non-line-of-sight (NLoS) scenarios \cite{b14}. To address this, reconfigurable intelligent surfaces (RIS) have been introduced to enhance the communication and sensing performance in ISAC systems \cite{b22}.

RIS is a two-dimensional array of phase shifters, adjusts reflection characteristics to optimize signal propagation and create virtual line-of-sight (VLoS) links in blocked paths \cite{b24}. Although RIS cannot decode signals, their phase configuration mitigates multipath effects, reduces attenuation, and enhances system performance. When integrated into ISAC systems, RIS improves both communication quality and radar accuracy, particularly in scenarios with weak or blocked line-of-sight (LoS) links \cite{b26}. RIS has been utilized in secure ISAC systems for downlink transmission \cite{b35}, wireless information and power transfer \cite{b36}, and in RIS-enhanced MISO systems to support single-antenna users \cite{b37}.

Most existing work assumes known channel state information (CSI) for all RIS-related links \cite{b22}, \cite{b35}. In practice, updating CSI each coherence interval incurs significant overhead, which grows with the number of RIS elements. To address this, bypassing channel estimation and designing passive beamforming based on user feedback has been explored. To the best of our knowledge, there is few works on feedback based RIS phase design. The author in \cite{b39} proposed a feedback-based beam training method using particle swarm optimization.

Motivated by the above facts, this paper considers the beamforming design in RIS-assisted mmWave ISAC systems with unknown CSI, where RIS provides VLoS links for targets in DFBS blind spots and users with severe path loss. Target detection in ISAC systems aims to overcome coverage interruptions caused by terrain, obstructions, and interference, ensuring continuous radar functionality. We also consider echo interference from multipath communication links during radar optimization. Various intelligent optimization algorithms are applied to the feedback-based beam training scheme to evaluate performance and stability. Our main contributions are summarized as follows:

\begin{itemize}
\item{We investigate the impact of echo interference from multipath communication links on sensing performance in RIS-assisted mmWave ISAC systems and propose a feedback-based beam training framework that enhances both sensing and communication performance.}

\item{Considering the high cost of cascaded channel estimation due to unknown CSI and the limited availability of feedback-based algorithms, we introduce a novel joint active and passive beam training scheme based on an improved artificial fish swarm algorithm (AFSA), achieving rapid convergence to the optimal solution without CSI.}

\item{Simulation results show that the feedback-based AFSA beam training scheme outperforms other intelligent optimization algorithms, demonstrating superior joint sensing and communication performance. Even with variations in user SNR and an expanded search domain, ASFA maintains stable performance.}

\end{itemize}

\section{System Model}
Consider an ISAC system with RIS, where a DFBS communicates with a single-antenna user while sensing a target in the far field of both DFBS and RIS. The DFBS, equipped with a uniform linear array (ULA) of $M$ antennas, transmits a unified ISAC waveform. The target and user are spatially separated, with no LoS link between DFBS and the target, and a direct but path-loss-affected link between DFBS and the user, as shown in Fig. 1.

In practical scenarios, the channel varies due to environmental factors. To maintain generality, we divide the beam training and transmission into two periods: $T_1$ for beam training and $T_2$ for ISAC signal transmission. The beam training period is split into $K$ sub-blocks, each with $S = \frac{T_1}{K}$ time slots, while the ISAC period is divided into $L$ sub-blocks, each with $U = \frac{T_2}{L}$ time slots. During the beam training period, the DFBS transmits ISAC signals and gradually updates the transmit beam at the DFBS and the reflection beam at the RIS based on echo superimposed signals and user feedback. During the ISAC period, the DFBS uses the trained beams to support both communication and sensing functions.

\vspace{-0.2cm}
\subsection{Downlink Transmit Signal Model}
During the beam training period, the DFBS transmits pilot signals to the user, who subsequently computes the received signal strength and provides feedback to the DFBS.

Let $\bm{w}(t) = [w_1(t),...,w_M(t)]^T \in \mathbb{C}^{M\times 1}$ denotes the continuous-time active beamforming vector. The transmitted pilot symbol $s(t)$ is assumed to be independent random variable with zero mean and unit covariance, then the transmitted pilot signal can be expressed as $\bm{x}(t) = \bm{w}(t)s(t)$.

\vspace{-0.2cm}
\subsection{RIS Model}
We model the RIS as a collection of discrete passive phase shifters, which can be individually controlled from the DFBS via a low-rate control link. Specifically, we assume that the RIS comprises $N$ elements and model its spatial response as that of a uniform rectangular array (URA). Let $\phi_i\in \left(0,2\pi\right]$ denote the phase shift of reflecting element $i\in\{1,2,\dots,N\}$ at the RIS, and the corresponding reflect beamforming matrix is denoted by $\bm{\Phi}=\text{diag}([e^{j\phi_1},\dots,e^{j\phi_N}])$. Let $\bm{\xi} = [e^{j\phi_1},\dots,e^{j\phi_N}]^T$, then $\bm{\Phi} = \text{diag}(\bm{\xi})$.

\vspace{-0.2cm}
\subsection{Communication and Radar Channel Model}
\begin{figure}[!t]
	\centering
	\includegraphics[width=2.5in]{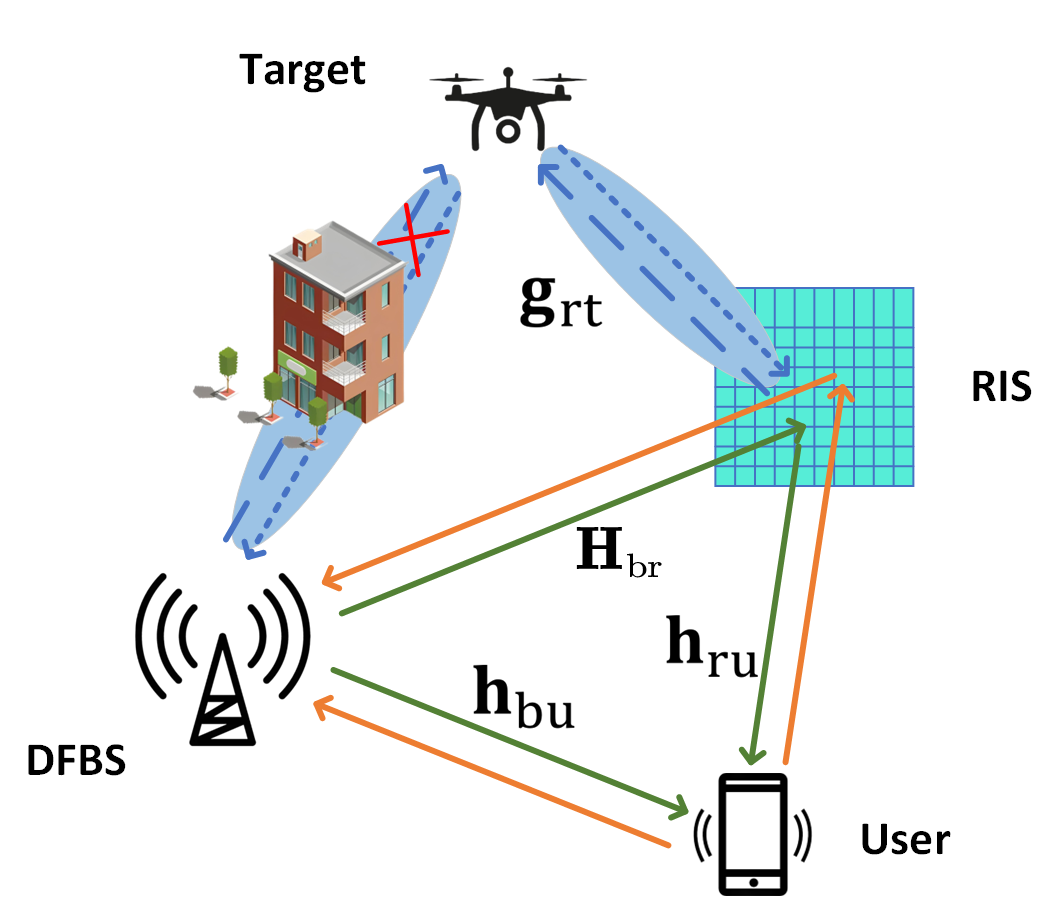}
	\caption{ISAC systems.}
	\vspace{-10pt} 
	\label{fig1}
\end{figure}

Let $\bm{H}_{br} \in \mathbb{C}^{N\times M}$ denote the DFBS-RIS channel matrix. The MISO channel vectors for the DFBS-User and RIS-User links are $\bm{h}_{bu} \in \mathbb{C}^{M\times 1}$ and $\bm{h}_{ru} \in \mathbb{C}^{N\times 1}$, respectively. The overall channel vector $\bm{h}_u^H$, including both the direct and RIS-assisted links, is given by:
\begin{align}
	\label{eq1}
	\bm{h}^H_u &=\bm{h}^H_{bu} + \bm{h}^H_{ru}\text{diag}(\bm{\xi})\bm{H}_{br} \nonumber \\
	&= \bm{h}^H_{bu} + \bm{\xi}^T\text{diag}(\bm{h}^H_{ru})\bm{H}_{br}.
\end{align}
Then the received signal of the user at time slot $t$ can be expressed as
\begin{align}
	\label{eq2}
	y_c(t) = \bm{h}^H_u \bm{w}(t)s(t) + n_c(t),
\end{align}
where $n_c(t)$ is the additive white Gaussian noise (AWGN) having zero mean and variance $\sigma^2_c$.

For the radar model, let $\bm{g}_{rt}\in\mathbb{C}^{N \times 1}$ denote the channel from the RIS to the target. The DFBS receives echo signals from the target and user through five paths: DFBS-RIS-target-RIS-DFBS, DFBS-RIS-user-RIS-DFBS, DFBS-RIS-user-DFBS, DFBS-user-DFBS, and DFBS-user-RIS-DFBS. The received superimposed echo signals at the DFBS can be expressed as:
\begin{align}
	\label{eq3}
	\bm{y}_s(t) &=  \rho_1(\text{diag}(\bm{g}^H_{rt})\bm{H}_{br})^H\bm{\xi}^*\bm{\xi}^T\text{diag}(\bm{g}^H_{rt})\bm{H}_{br}\bm{w}(t)s(t) \nonumber \\ &+[\rho_2(\text{diag}(\bm{h}^H_{ru})\bm{H}_{br})^H\bm{\xi}^*\bm{\xi}^T\text{diag}(\bm{h}^H_{ru})\bm{H}_{br} \nonumber \\
	&+\rho_3\bm{h}_{bu}\bm{\xi}^T\text{diag}(\bm{h}^H_{ru})\bm{H}_{br} + \rho_4\bm{h}_{bu}\bm{h}^H_{bu} \nonumber \\ &+\rho_5(\text{diag}(\bm{h}^H_{ru})\bm{H}_{br})^H\bm{\xi}^*\bm{h}^H_{bu}]\bm{w}(t)s(t) + n_s(t),
\end{align}
where $\rho_1$, $\rho_2$, $\rho_3$, $\rho_4$ and $\rho_5$ denote the echo coefficients from RIS to RIS via target, from RIS to RIS via user, from RIS to DFBS via user, from DFBS to DFBS via user, and from DFBS to RIS via user, respectively.

\vspace{-0.2cm}
\subsection{ISAC Period}
In the ISAC period, the trained active and passive beamforming vectors, $\bm{w}_{\mathrm{opt}}(t)$ and $\bm{\xi}_{\mathrm{opt}}(t)$, are sent to the DFBS and RIS for transmitting ISAC waveforms. The DFBS transmits the precoded signal $\tilde{\bm{x}}(t) = \bm{w}_{\mathrm{opt}}(t)d(t)$ during time slots $t \in\{T_1 + 1, \dots, T_1 + T_2\}$, where $d(t)$ is the communication symbol. Similarly, the echo signal $\tilde{\bm{y}}_s(t)$ at the DFBS and the received signal $\tilde{\bm{y}}_c(t)$ at the user follow the same form.

\section{Joint Active and Passive \\ Beam Training Based on AFSA}
\subsection{Communication and radar sensing metrics}
In a single-user MISO communication with the DFBS, the service quality is determined by the SINR at the user. Since integrated signals are used, the communication user is unaffected by radar signals. Therefore, SNR is used as the performance indicator instead of SINR.

According to the (\ref{eq1}) and (\ref{eq2}), the SNR $\gamma$ at the user can be expressed as
\begin{equation}
	\label{eq4}
	\gamma = \frac{|\bm{h}^H_u\bm{w}(t)|^2}{\sigma^2_c} 
	= \frac{\bm{h}^H_u\bm{w}(t)\bm{w}^H(t)\bm{h}_u}{\sigma^2_c}.
\end{equation}
Since the only interference in (4) is the noise term, and AWGN is considered constant over time, the SNR metric can be expressed as the total power $|y_c(t)|^2$ of the received signal at the user, given by:
\begin{align}
	\label{eq5}
	|y_c(t)|^2 &= |\bm{h}^H_u\bm{w}(t)|^2 + \sigma^2_c \nonumber \\
	&= \bm{h}^H_u\bm{w}(t)\bm{w}^H(t)\bm{h}_u + \sigma^2_c.
\end{align}

An alternative communication performance metric is the rate of the user, which plays the same role for an single user. Hence, we adopt $|y_c(t)|^2$ as the communication performance metric.

Due to the lack of CSI for the user and target, the received echo signal power $|y_s(t)|^2$ is used as the evaluation metric for sensing performance in beamforming design. Unlike the user’s received signal power, the echo signals at the DFBS also include interference from multiple paths: DFBS-RIS-user-RIS-DFBS, DFBS-RIS-user-DFBS, DFBS-user-DFBS, and DFBS-user-RIS-DFBS. Thus, radar performance optimization must address the interference from the user. A specific solution will be provided in the next subsection.

\vspace{-0.2cm}
\subsection{Design problem}
For the DFBS echo superimposed signal, it is noted that the interference echo from the user is a destructive interference signal. Therefore, to limit the interference signal power increase caused by the growth of $|y_c(t)|^2$, we strictly constrain the user's received signal power to satisfy constraint $|y_c(t)|^2 \leq \eta_{\mathrm{max}}$. Additionally, to ensure basic communication requirements, constraint $|y_c(t)|^2 \geq \eta_{\mathrm{min}} $ must also be satisfied. Accordingly, we formulate the following optimization problem
\begin{subequations}\label{eq6}
	\begin{align}
		(\text{P}):\: \mathop{\max}_{\bm{w}(t),\bm{\xi}(t)} \ & \parallel \bm{y}_s(t) \parallel^2 \nonumber \\ 
		s.t. \ \ & |y_c(t)|^2 \in [\eta_{\mathrm{min}},\eta_{\mathrm{max}}], \label{eq6a} \\ 
		& \parallel \bm{w}(t) \parallel^2 = P, \label{eq6b} \\
		& |\xi_n(t)|=1,n\in\{1,\dots,N\}. \label{eq6c}
	\end{align}
\end{subequations}
where $P$ denotes the DFBS transmit power, (\ref{eq6c}) is the RIS phase shift constraint, and $\eta_{\mathrm{max}}$ and $\eta_{\mathrm{min}}$ are the maximum and minimum received signal power thresholds at the user. The above problem $(\text{P})$ is non-convex. Traditional methods cannot be used without CSI, so we propose a feedback beam training scheme based on AFSA \cite{b40}.

\subsection{Algorithm initialization}
During the beamforming training period, at the start of sub-block $l=1$, $S$ active and passive beam pairs $\{\bm{w}(t), \bm{\xi}(t), t \in {1, \dots, S}\}$ are randomly generated. For subsequent sub-blocks $l \in \{2, \dots, K\}$, The feasibility of the beam depends on whether the result from the previous sub-block satisfies the constraint (6a).

We consider three cases based on beam feasibility and propose corresponding training methods. In case A, where all beams are feasible, we use a modified AFSA for beam training. In cases B and C, where some or all beams are infeasible, we adjust the AFSA update formulas to move infeasible beams toward the feasible range in (\ref{eq6a}). The designed $S$ active and passive beam pairs for sub-block $l$ are assigned to the DFBS and RIS according to the previous sub-block’s results.

AFSA is an optimization algorithm inspired by the collective behavior of fish. Each artificial fish (AF) represents a potential solution, and its movement in the search space adjusts the solution parameters. The AF swarm set is denoted as $\mathcal{S}^{(l)} = \{1,2,\dots,S\}$, with $S$ beam fishes at the $l$-th sub-block. The position of the $i$-th AF at the $l$-th sub-block is $\bm{X}^{(l)}_i, i=\{1,2,\dots,S\}$, which contains $M+N$ elements, i.e.,
\begin{equation}
	\label{eq7}
	\bm{X}^{(l)}_i = [(\bm{w}^{(l)}_i)^T, (\bm{\zeta}^{(l)}_i)^T],i\in\mathcal{S}^{(l)},l\in\{1,\dots,K\},
\end{equation}
where $\bm{w}^{(l)}_i = [w^{(l)}_{i,1},\dots,w^{(l)}_{i,M}]^T$, $\bm{\zeta}^{(l)}_i = [\phi^{(l)}_{i,1},\dots,\phi^{(l)}_{i,N}]^T$ denote the potential solutions provided by the $i$-th AF at sub-block $l$ for the active beamforming vector and RIS phase shifts, respectively. Specifically, for the $m$-th element in $\bm{w}^{(l)}_i$, $w^{(l)}_{i,m}$ can be expressed as
\begin{equation}
	\label{eq8}
	w^{(l)}_{i,m} = \beta^{(l)}_{i,m} \times e^{j\theta^{(i)}_{i,m}}, m\in\{1,2,\dots,M\},
\end{equation}
where $\beta^{(l)}_{i,m} \in [0,\sqrt{P}]$ is the amplitude coefficient and $\theta^{(i)}_{i,m} \in [-\pi,\pi)$ is the phase shift coefficient with the transmit power constraint $\parallel\bm{w}(t)\parallel^2 = P$.
In addition, for the $n$-th element in $\bm{\zeta}^{(l)}_i$, $\phi^{(l)}_{i,n}\in[0,2\pi)$.

\subsection{Beam Training Methods Based on AFSA for Each Sub-Block}
We define the fitness function of the AF swarm as follows. For AF $i^{(l)}$ with position $\bm{X}^{(l)}_i$, the fitness function is given by
\begin{equation}
	\label{eq9}
	F(\bm{X}^{(l)}_i) = \parallel \bm{y}_s(\bm{w}^{(l)}_i,\bm{\zeta}^{(l)}_i)\parallel^2.
\end{equation}

Next, we define an auxiliary function to measure the communication performance of beam AF $i^{(l)}$ with respect to constraint (\ref{eq6a})
\begin{equation}
	\label{eq10}
	f(\bm{X}^{(l)}_i) = |\bm{y}_c(\bm{w}^{(l)}_i,\bm{\zeta}^{(l)}_i)|^2
\end{equation}
Based on the auxiliary function (\ref{eq10}), we place all feasible AFs into a feasible AF set
\begin{equation}
	\label{eq11}
	R^{(l)} = \{i^{(l)}|f(\bm{X}^{(l)}_i)\in[\eta_{\mathrm{min}},\eta_{\mathrm{max}}]\}.
\end{equation}
We use $\bm{X}^{(l-1)}_{\mathrm{opt}}$ to denote the globally optimal position up to sub-block $l-1$, and $F(\bm{X}^{(l-1)}_{\mathrm{opt}})$ for the corresponding global optimal fitness.

There are three cases depending on the feasibility of the AFs from sub-block $l-1$. For each case, we propose corresponding beam training methods, summarized in Algorithm 1.

\textit{1}) Case A: All the $S$ beam AFs of the sub-block $l-1$ are feasible, i.e., $R^{(l-1)}=S, l\in\{2,\dots,K\}$. 

In this case, we introduce $\chi_1$ and $\chi_2$ as the active and passive steps of beam pairs, and $\nu_1$ and $\nu_2$ as the active and passive visions. The modified AFSA employs these parameters in the beam training process, adjusting the active and passive beamforming vectors based on their complex and real domain differences and value ranges. 

In the process of using the modified ASFA to solve the beam training problem, the three basic behaviors of the modified AFSA are mainly employed as follows:

\newcounter{TempEqCnt}
\setcounter{TempEqCnt}{\value{equation}}
\setcounter{equation}{24}
\begin{figure*}[ht]
	\centering
	\begin{equation}
		\label{eq25}
		\bar{i}^{(l-1)} = 
		\begin{cases}
			\mathrm{argmax}_{i^{(l-1)}\in R^{(l-1)}} \mbox{ } f(\bm{X}^{(l-1)}_{i}), &{\text{if}}\ \mathrm{max} \mbox{ } f(\bm{X}^{(l-1)}_{i}) < \eta_{\mathrm{min}}\\
			\mathrm{argmin}_{i^{(l-1)}\in R^{(l-1)}} \mbox{ } f(\bm{X}^{(l-1)}_{i}), &{\text{if}}\ \mathrm{min} \mbox{ } f(\bm{X}^{(l-1)}_{i}) > \eta_{\mathrm{max}}\\
			\mathrm{argmin}_{i^{(l-1)}\in R^{(l-1)}} \mbox{ } \mathrm{min}\{|f(\bm{X}^{(l-1)}_{i})-\eta_{\mathrm{min}}|,|f(\bm{X}^{(l-1)}_{i})-\eta_{\mathrm{max}}|\}, &{\text{otherwise.}}
		\end{cases}		
	\end{equation}
	\vspace*{8pt}
	\hrulefill
\end{figure*}
\setcounter{equation}{\value{TempEqCnt}}

(\textit{a}) Foraging behavior: For the $i$-th AF with position $\bm{X}^{(l-1)}_i = [\bm{w}^{(l-1)}_i, \bm{\zeta}^{(l-1)}_i]$, we treat $\bm{w}^{(l-1)}_i$ and $\bm{\zeta}^{(l-1)}_i$ as two separate active and passive AFs. The new visual positions are $\bm{w}^{(l)}_{i,e}$ and $\bm{\zeta}^{(l)}_{i,e}$, respectively. If $F(\bm{X}^{(l)}_{i,e}) > F(\bm{X}^{(l-1)}_i)$, the AF moves one step towards $\bm{X}^{(l)}_{i} = [\bm{w}^{(l)}_i, \bm{\zeta}^{(l)}_i]$. The process is as follows:
\begin{equation}
	\label{eq12}
	\bm{w}^{(l)}_{i,e} = \bm{w}^{(l-1)}_{i} + \frac{\nu_1}{\sigma_1}  \cdot Rand(),
\end{equation}
\begin{equation}
	\label{eq13}
	\bm{\zeta}^{(l)}_{i,e} = \bm{\zeta}^{(l-1)}_{i} + \frac{\nu_2}{\sigma_2} \cdot Rand(),
\end{equation}
\begin{equation}
	\label{eq14}
	\bm{w}^{(l)}_{i} = \bm{w}^{(l-1)}_{i} + \frac{\bm{w}^{(l)}_{i,e}-\bm{w}^{(l-1)}_{i}}{\parallel\bm{w}^{(l)}_{i,e}-\bm{w}^{(l-1)}_{i}\parallel} \cdot \chi_1 \cdot r_1,
\end{equation}
\begin{equation}
	\label{eq15}
	\bm{\zeta}^{(l)}_{i} = \bm{\zeta}^{(l-1)}_{i} + \frac{\bm{\zeta}^{(l)}_{i,e}-\bm{\zeta}^{(l-1)}_{i}}{\parallel\bm{\zeta}^{(l)}_{i,e}-\bm{\zeta}^{(l-1)}_{i}\parallel} \cdot \chi_2 \cdot r_2.
\end{equation}
In the equations, $Rand()$ is a random number in $(-1,1)$, and $r_1$ and $r_2$ are in $(0,1)$. Due to the large geometric distance between high-dimensional vectors, larger AF vision is needed, with correction factors $\sigma_1$ and $\sigma_2$ introduced.

If $F(\bm{X}^{(l)}_{i,e}) \leq F(\bm{X}^{(l-1)}_{i})$, AFs will search a new position. If searches exceed $T_{\mathrm{max}}$, AFs will randomly move one step
\begin{equation}
	\label{eq16}
	\bm{w}^{(l)}_{i} = \bm{w}^{(l-1)}_{i} + \frac{\nu_1}{\sigma_1} \cdot Rand(),
\end{equation}
\begin{equation}
	\label{eq17}
	\bm{\zeta}^{(l)}_{i} = \bm{\zeta}^{(l-1)}_{i} + \frac{\nu_2}{\sigma_2} \cdot Rand(),
\end{equation}

For simplicity, we refer to the active and passive steps $\chi_1$ and $\chi_2$ as $\chi$, and the active and passive visions $\nu_1$ and $\nu_2$ as $\nu$, though they are still processed separately in practice.

(\textit{b}) Clustering behavior: given $\bm{X}^{(l-1)}_{i}$, AF $i^{(l-1)}$ explores $n_f$ partners ($d^{(l-1)}_{ij} < \nu$). Let $\bm{X}^{(l-1)}_{i,c}$ be the center of these $n_f$ AFs, and $F(\bm{X}^{(l-1)}_{i,c})$ the fitness. If $F(\bm{X}^{(l-1)}_{i,c})/n_f > \delta F(\bm{X}^{(l-1)}_{i})$, AF $i^{(l-1)}$ updates its position using the formula
\begin{equation}
	\label{eq18}
	\bm{X}^{(l)}_{i} = \bm{X}^{(l-1)}_{i} + \frac{\bm{X}^{(l-1)}_{i,c} - \bm{X}^{(l-1)}_{i}}{\parallel \bm{X}^{(l-1)}_{i,c} - \bm{X}^{(l-1)}_{i}\parallel} \cdot \chi \cdot r.
\end{equation}
where $r$ is a random number in $(0,1)$. Otherwise, foraging behavior is performed.

(\textit{c}) Rear chasing behavior: given $\bm{X}^{(l-1)}_{i}$, AFs search for $n_f$ partners in the current field ($d^{(l-1)}_{ij} < \nu$), where $\bm{X}^{(l-1)}_{i,\mathrm{max}}$ has the highest fitness. If $F(\bm{X}^{(l-1)}_{i,\mathrm{max}})/n_f > \delta F(\bm{X}^{(l-1)}_{i})$, AFs updates its position as:
\begin{equation}
	\label{eq19}
	\bm{X}^{(l)}_{i} = \bm{X}^{(l-1)}_{i} + \frac{\bm{X}^{(l-1)}_{i,\mathrm{max}} - \bm{X}^{(l-1)}_{i}}{\parallel \bm{X}^{(l-1)}_{i,\mathrm{max}} - \bm{X}^{(l-1)}_{i}\parallel} \cdot \chi \cdot r.
\end{equation}
Otherwise, foraging behavior will be carried out.

The execution order is as follows: Before updating the position, the AF $i^{(l-1)}$ compares the fitness of the new positions from clustering and rear chasing behaviors, and chooses the one with higher fitness.

However, once the updated position is not within bounds, the amendments
\begin{align}
	\label{eq20}
	&|w_{i,m}| = \text{max}\{z_{\text{lower}},|w_{i,m}|\}, \ \ \ |w_{i,m}| = \text{min}\{z_{\text{upper}},|w_{i,m}|\}, \nonumber \\
	&\zeta_{i,n} = \text{max}\{\zeta_{\text{lower}},\zeta_{i,n}\}, \ \ \ \ \ \ \ \ \ \ \ \ \ \ \ \zeta_{i,n} = \text{min}\{\zeta_{\text{upper}},\zeta_{i,n}\},
\end{align}
will be carried out, where $z_{\text{upper}} = \sqrt{P}$, $z_{\text{lower}} = 0$ are the bounds for the active beam amplitudes, and $\zeta_{\text{upper}} = 2\pi$, $\zeta_{\text{lower}} = 0$ are the bounds for the passive beam phase.

Based on the update formulas (12)-(19), we can obtain $S$ new position $\{\bm{X}^{(l)}_i\}$ for $S$ feasible AF of the sub-block $l$.

\textit{2}) Case $B$: Partial beam AFs of the sub-block $l-1$ are feasible, i.e., $0 < R^{(l-1)} < S, l\in\{2,\dots,K\}$.

In this case, the infeasible AF moves towards the best performing target AF (if $R^{(l-1)} > 1$) to generate a new beam AF. The target AF is determined by the following formula.
\begin{equation}
	\label{eq21}
	\bar{i}^{(l-1)} =
	\begin{cases}
		i^{(l-1)} \in R^{(l-1)}, &{\text{if}}\ R^{(l-1)}=1, \\
		\mathrm{argmax}_{i^{(l-1)} \in R^{(l-1)}} F(\bm{X}^{(l-1)}_i), &{\text{if}}\ R^{(l-1)} > 1,
	\end{cases}
\end{equation}
Then these infeasible AFs update their positions based on the position of the target AF $\bar{i}^{(l-1)}$, expressed as
\begin{equation}
	\label{eq22}
	\bm{X}^{(l)}_{i} = \bm{X}^{(l-1)}_{i} + \frac{\bm{X}^{(l-1)}_{\bar{i}}-\bm{X}^{(l-1)}_{i}}{\parallel \bm{X}^{(l-1)}_{\bar{i}}-\bm{X}^{(l-1)}_{i}\parallel} \cdot \chi \cdot r
\end{equation}

The remaining feasible AFs will perform the three basic behaviors as in Case A (Eqs. (\ref{eq12})-(\ref{eq19})), while infeasible AFs will be excluded. The global optimal solution is updated for sub-block $l-1$ as follows.
\begin{equation}
	\label{eq23}
	\bm{X}^{(l-1)}_{\mathrm{opt}} = 
	\begin{cases}
		\bm{X}^{(l-2)}_{\mathrm{opt}}, &{\text{if}}\ F(\bm{X}^{(l-1)}_{\bar{i}}) < F(\bm{X}^{(l-2)}_{\mathrm{opt}}),\\
		\bm{X}^{(l-1)}_{\bar{i}}, &{\text{if}}\ F(\bm{X}^{(l-1)}_{\bar{i}}) \geq F(\bm{X}^{(l-2)}_{\mathrm{opt}}),
	\end{cases}
\end{equation}
Accordingly, the global fitness becomes
\begin{equation}
	\label{eq24}
	F(\bm{X}_{\mathrm{opt}}^{(l-1)}) = 
	\begin{cases}
		F(\bm{X}_{\mathrm{opt}}^{(l-2)}), &{\text{if}}\ F(\bm{X}^{(l-1)}_{\bar{i}}) < F(\bm{X}^{(l-2)}_{\mathrm{opt}}),\\
		F(\bm{X}^{(l-1)}_{\bar{i}}), &{\text{if}}\ F(\bm{X}^{(l-1)}_{\bar{i}}) \geq F(\bm{X}^{(l-2)}_{\mathrm{opt}}),
	\end{cases}
\end{equation}

Finally, with the above steps, we can obtain $S$ new positions in sub-block $l$ with respect to $S$ AFs.

\textit{3}) Case C: All the beam AFs of the sub-block $l-1$ are infeasible, i.e., $R^{(l-1)}=0, l\in\{2,\dots,K\}$.

In this case, infeasible AFs gradually move towards the communication feasible interval $[\eta_{\mathrm{min}}, \eta_{\mathrm{max}}]$. The AFs (except the target AF) update towards the target AF.

We calculate the received communication power $f(\bm{X}^{(l-1)}_{i})$ for all $S$ beam AFs, sort them, and identify the target AF closest to the communication interval. The specific method is
shown in Eq. (\ref{eq25}) at the top. Infeasible AFs then update their positions based on the target AF's position using Eq. (\ref{eq22}). The target AF uses Eq. (\ref{eq12})-(\ref{eq17}) for foraging behavior to generate a new AF.

Since all AFs in case C are infeasible, no global optimal update occurs, and the value of $F$ dose not get worse.

\subsection{Convergence and complexity analysis}
We propose a convergent active and passive beam training algorithm based on an improved ASFA method. For sub-block $l$, the global optimal position $\bm{X}^{(l)}_{\mathrm{opt}}$ is found, and using Eq. (9), we obtain the maximum echo power $F(\bm{X}^{(l)}_{\mathrm{opt}})$, related to the maximum sensing echo power from the previous sub-block
\setcounter{equation}{25}
\begin{equation}
	\label{eq26}
	F(\bm{X}^{(l)}_{\mathrm{opt}}) \geq F(\bm{X}^{(l-1)}_{\mathrm{opt}}),l>1,
\end{equation}
The maximum echo signal power never decreases. AFs either perform the three basic behaviors or moves toward the target position, ensuring convergence. The modified behavior ensures AFs move toward the global optimum while satisfying communication constraints, with an upper limit on the objective function. Thus, Algorithm \ref{alg1} converges.

The complexity of the proposed algorithm is $\mathcal{O}((K-1)(4R+R \times T_{\mathrm{max}})(M+N))$, where $K-1$ is the number of iterations, $R$ is the number of beam AFs, $T_{\mathrm{max}}$ is the maximum number of attempts in the foraging behavior, and $M+N$ is the dimension of each AF.

\begin{algorithm}[t]
	\caption{Beam Training Method Based on ASFA.}
	\label{alg1}
	\begin{algorithmic}[1]
		\renewcommand{\algorithmicrequire}{\textbf{Input:}}
		\renewcommand{\algorithmicensure}{\textbf{Output:}}                                                              
		\REQUIRE $S^{(l-1)}$, $\{\bm{X}^{(l-1)}_{i}\}$, $P$, $\chi_1$, $\chi_2$, $\nu_1$, $\nu_2$                 
		\ENSURE $\bm{X}^{(l-1)}_{\mathrm{opt}}$, $F(\bm{X}^{(l-1)}_{\mathrm{opt}})$, $\{\bm{X}^{(l)}_{i},i^{(l)} \in S^{(l)} \}$       
		\STATE Calculate fesaible set $R^{(l-1)}$.
		\IF {$R^{(l-1)} = S$}
		\STATE Perform clustering behavior or foraging behavior to obtain $\bm{X}^{(l)}_{i,1}$ and $F(\bm{X}^{(l)}_{i,1})$ by (\ref{eq9}), (\ref{eq12})-(\ref{eq17}), (\ref{eq18}).
		\STATE Perform chasing behavior or foraging behavior to obtain $\bm{X}^{(l)}_{i,2}$ and $F(\bm{X}^{(l)}_{i,2})$ by (\ref{eq12})-(\ref{eq17}), (\ref{eq19}).
		\STATE $\bm{X}^{(l)}_{i} = \text{argmax}\{F(\bm{X}^{(l)}_{i,1}), F(\bm{X}^{(l)}_{i,2})\}$
		\ELSIF {$0 < R^{(l-1)} < S$}
		\STATE Select the target AF $\bar{i}^{(l-1)}$ by using (\ref{eq21}) and set the target position $\bm{X}^{(l-1)}_{\bar{i}}$
		\STATE Update the position $\{\bm{X}^{(l)}_i\}$ and obtain $\bm{X}^{(l-1)}_{\mathrm{opt}}$ by using (\ref{eq12})-(\ref{eq19}), (\ref{eq22}), (\ref{eq23})
		\ELSE
		\STATE Select the target AF $\bar{i}^{(l-1)}$ by using (\ref{eq25}) and set the target position $\bm{X}^{(l-1)}_{\bar{i}}$
		\STATE Update the position $\{\bm{X}^{(l)}_i\}$ and obtain $\bm{X}^{(l-1)}_{\mathrm{opt}}$ by using (\ref{eq12})-(\ref{eq17}), (\ref{eq23}), (\ref{eq25})
		\ENDIF
	\end{algorithmic}
\end{algorithm}

\section{Simulation Results}
This section presents numerical results to evaluate the proposed feedback-based beam training scheme using AFSA for the RIS-enabled ISAC system. The locations of the DFBS, RIS, target, and user are (0m, 0m, 15m), (30m, 0m, 10m), (15m, -25m, 0m), and (15m, 30m, 0m), respectively. In the simulation, we compare the training results with those obtained using the Particle Swarm Optimization (PSO) \cite{b39} algorithm and the Ant Colony Optimization (ACO) \cite{b41} algorithm.

\begin{figure}[!t]
	\centering
	\includegraphics[width=2.8in]{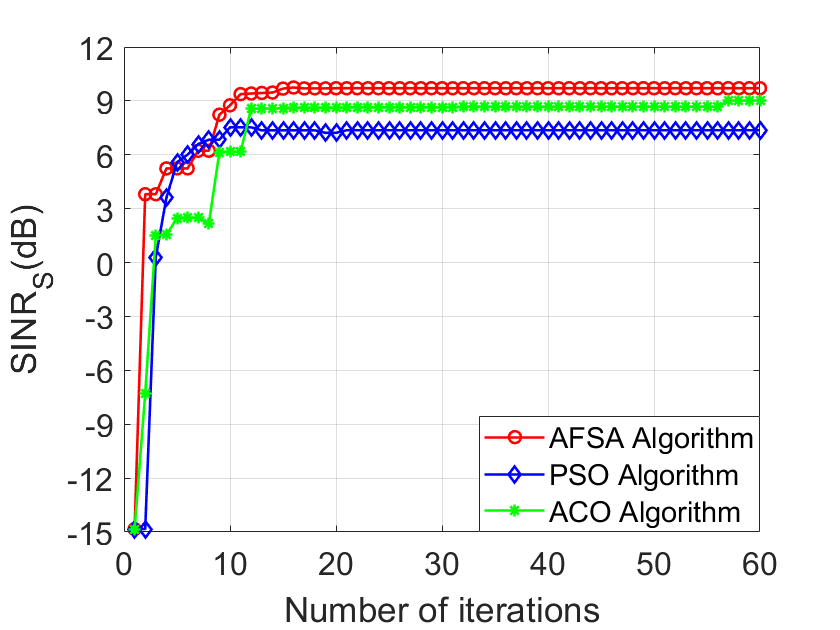}
	\caption{The convergence performance of the proposed AFSA-based beam training scheme, compared to PSO and ACO-based schemes, when $M = 8$ and $N = 64$.}
	\label{fig4}
\end{figure}

Fig. \ref{fig4} shows the convergence performance of the AFSA, PSO, and ACO-based beam training schemes. The AFSA-based scheme converges to a stable solution within 20 iterations, outperforming the other two. In contrast, the PSO algorithm experiences fluctuations at the 50th iteration, indicating poor convergence. The ACO algorithm performs better than PSO but still lags behind AFSA in both convergence speed and final performance.

\begin{figure}[!t]
	\centering
	\includegraphics[width=2.8in]{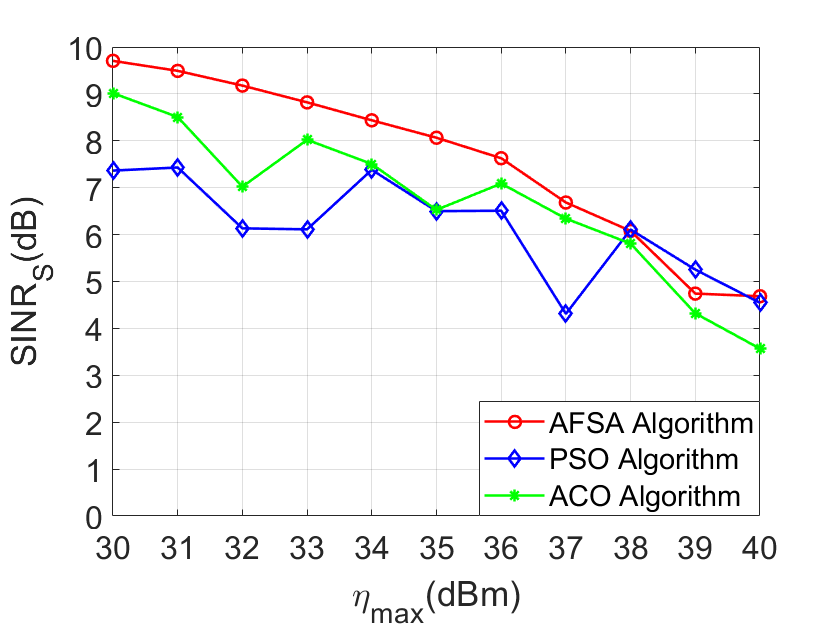}
	\caption{Sensing performance of the three algorithms versus various communication upper-bound constraints $\eta_{\mathrm{max}}$ when $M = 8$, $N = 64$.}
	\label{fig5}
\end{figure}

Fig. \ref{fig5} shows the sensing performance of the ISAC system with the proposed improved AFSA and comparison algorithms under various user received power constraints. As the upper-bound constraints loosen, communication constraints become relaxed, leading to performance degradation for all algorithms. However, the AFSA consistently performs near-optimal, while the PSO algorithm is generally worse, with occasional slight improvements. The ACO algorithm shows the largest performance decline and instability, confirming the superiority of the improved AFSA in both performance and stability.

\begin{figure}[!t]
	\centering
	\includegraphics[width=2.8in]{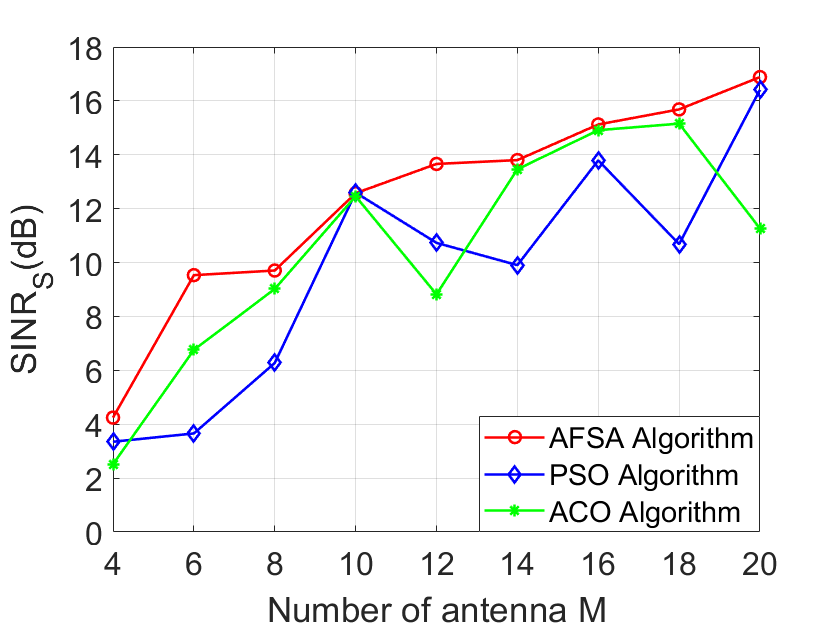}
	\caption{Sensing performance of the three algorithms versus the number of transmit antennas $M$ at the DFBS when $N=64$.}
	\label{fig7}
\end{figure}

Fig. \ref{fig7} shows the sensing performance of the three algorithms as the transmitting antenna size $M$ increases, while $N$ is fixed at $64$. The AFSA shows significant improvement in performance due to increased active beamforming gain. In contrast, the PSO and ACO algorithms exhibit large volatility, as the higher antenna count increases the search dimension, demanding more from the algorithms’ search capabilities.

\begin{figure}[!t]
	\centering
	\includegraphics[width=2.8in]{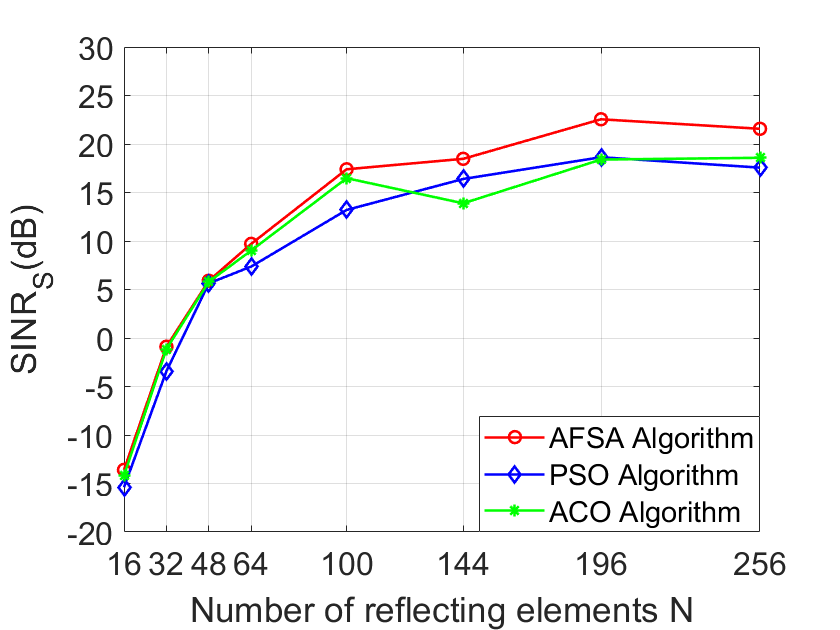}
	\caption{Sensing performance of the three algorithms versus the number of reflecting elements $N$ at the RIS when $M=8$.}
	\label{fig9}
\end{figure}

Fig. \ref{fig9} shows the sensing performance of the three algorithms as the number of reflective elements $N$ increases, while $M$ is fixed at $8$. As $N$ grows, performance improves due to higher beamforming and aperture gains. The AFSA achieves the highest gain, followed by ACO, with PSO performing the worst.

The AFSA outperforms the PSO and ACO algorithms in most cases, maintaining better stability in high-dimensional search spaces. This demonstrates the effectiveness of the proposed feedback-based beam training scheme.

\section{Conclusion}
In this paper, we propose a beamforming design for RIS-assisted ISAC systems, addressing the beam training problem to enhance sensing performance while ensuring communication quality, considering the interference from communication user echoes. We introduce a feedback-based joint active and passive beam training scheme using an improved AFSA algorithm, which improves both sensing and communication performance without the need for high overhead in channel estimation. Experimental results demonstrate that, despite the increased search space with more array antennas and RIS elements, the AFSA algorithm outperforms other algorithms like PSO and ACO in both search efficiency and stability.

\end{document}